\documentclass[reqno,10pt]{amsart}		

\calclayout

\usepackage[english]{babel}
\usepackage{amsfonts, amsmath, amssymb, amsthm}
\usepackage{hyperref}

\theoremstyle{remark}

\usepackage{algorithm}
\usepackage[noend]{algpseudocode}

\usepackage{braket}

\def\RR{\mathbb{R}}

\def\Bcal{\mathcal{B}}

\DeclareMathOperator{\Tr}{Tr}

\usepackage{multirow}
\usepackage{array}
    \newcolumntype{P}[1]{>{\centering\arraybackslash}p{#1}}

\title[]{Adaptive Pauli Shadows for Energy Estimation}

\usepackage[foot]{amsaddr}
\author{Charles Hadfield}
\email{charles.hadfield@ibm.com}

\address{IBM Quantum, IBM T.J. Watson Research Center, Yorktown Heights, NY 10598}

\begin{document}

\maketitle
\begin{abstract}
Locally-biased classical shadows allow rapid estimation of energies of quantum Hamiltonians. Recently, derandomised classical shadows have emerged claiming to be even more accurate. This accuracy comes at a cost of introducing classical computing resources into the energy estimation procedure. This present note shows, by adding a fraction of this classical computing resource to the locally-biased classical shadows setting, that the modified algorithm, termed \emph{Adaptive Pauli Shadows} is state-of-the-art for energy estimation.
\end{abstract}

\thispagestyle{empty}

\begin{section}{Introduction}
Techniques falling under the umbrella of classical shadows are in vogue within the quantum information community \cite{simons2021}. The tools of classical shadows have been attractively applied to tomographic problems \cite{aaronson2017, hkp2020}. In a recent work, with co-authors from IBM, we showed \cite{hadfield2020} how to adapt the tool of classical shadows to the different setting of energy estimation of a \emph{single} linear combination of Pauli observables. This is precisely the problem encountered when estimating the energy of a quantum Hamiltonian; a problem ubiquitous in quantum computing due to the use of variational quantum eigensolver algorithms. 
We called the technique \emph{locally-biased classical shadows}, or LBCS for short.
This algorithm is noise-resilient in the same way that the technique of classical shadows for tomographic problems is noise resilient \cite{noise2020}. Moreover, we claimed in a certain context, that our algorithm was state-of-the-art for the problem of energy estimation. Specifically we assumed a context where, after preparing the quantum state whose energy must be estimated, the quantum processor is not allowed to perform any entangling gates. To claim this, we benchmarked the algorithm on estimating the ground energy given the ground state of various Hamiltonians from quantum chemistry. (The Hamiltonians were mapped to quantum processors of size between 4 and 16 qubits.)

A technique of derandomising classical shadows has very recently been introduced and applied to the energy estimation problem. It claims to be state-of-the-art in the regime appropriate to LBCS. However it incurs a classical computational cost not present in the LBCS algorithm. If we allow LBCS access to a similar classical computation (actually a cost which is assymptotically better), then this modified algorithm performs similarly to the derandomising procedure, and in many cases, particularly as we move to larger instances, appears to be more accurate than the derandomising procedure. Let me call this new algorithm \emph{Adaptive Pauli Shadows}; it is the clear modification of LBCS that one would make if one were to allow oneself more classical compute resources.

Let us finish this introduction by considering the performance of four different algorithms for estimating the energy of a quantum Hamiltonian. The Hamiltonians are from small molecules: H\textsubscript{2}, LiH, BeH\textsubscript{2}, H\textsubscript{2}O, NH\textsubscript{3}. They are mapped onto 8, 12, 14, 14, 16 qubits respectively and all are mapped using the three common fermionic encodings: Jordan-Wigner, Parity, Bravyi-Kitaev. Given access to the true ground state $\rho$ and ground energy $E_G$ (in units of Hartree) we allow each of the four different algorithms to have access to $\rho$ 1000 times. Afterwards, each algorithm gives an estimate $\hat E_G$ and we record the average error of $|E_G-\hat E_G|$. The four algorithms are: 
\begin{itemize}
    \item Classical Shadows using random Pauli measurements \cite{hkp2020} (CS);
    \item Locally-biased classical shadows \cite{hadfield2020} (LBCS);
    \item Derandomized classical shadows \cite{hkp2021} (Derand.);
    \item Adaptive Pauli shadows (APS).
\end{itemize}
In order to compute the average error $|E_G-\hat E_G|$ for each algorithm we do the following: for CS and LBCS, the algorithms have a known variance, $\mathrm{var}$, calculated in \cite{hadfield2020} from which the average error is $\sqrt{\mathrm{var}/1000}$; for Derand.\ and APS, we simulate the process ten times and calculate a root-square-mean deviation as proposed in \cite{hkp2021}.
(It should be stated explicitly that CS was never intended for the problem of energy estimation of a single observable.)

\begin{table}[h]
\small
\begin{tabular}{|c | l ||  P{2cm} | P{2cm} | P{2cm} | P{2cm} |}
    \hline
        \multirow{2}{*}{\shortstack{Molecule\\($E_G$ in Hartree) (qubits)}}
        &
        \multirow{2}{*}{Encoding}
        &
        \multicolumn{4}{c|}{Average energy estimation error} \\
        \cline{3-6}
        &
        & {CS}
        & {LBCS}
        & {Derand.} 
        & {APS} \\
        \hline
        \hline
        \multirow{3}{*}{\shortstack{H$_2$\\(-1.86) \,\, (8)}}
        & JW
        & 0.23
        & 0.13
        & { 0.06}
        & 0.08
        \\
        & Parity
        & 0.27
        & 0.14
        & { 0.03}
        & 0.05
        \\
        & BK
        & 0.41
        & 0.14
        & { 0.06}
        & 0.08
        \\
        \hline
        \multirow{3}{*}{\shortstack{LiH\\(-8.91)\,\, (12)}}
        & JW
        & 0.52
        & 0.12
        & { 0.03}
        & 0.04
        \\
        & Parity
        & 0.87
        & 0.16
        & 0.03
        & 0.05
        \\
        & BK
        & 0.40
        & 0.26
        & 0.04
        & 0.07
        \\
        \hline
        \multirow{3}{*}{\shortstack{BeH$_2$\\(-19.0)\,\, (14)}}
        & JW
        & 1.29
        & 0.26
        & { 0.06}
        & { 0.06}
        \\
        & Parity
        & 1.78
        & 0.36
        & 0.09
        & { 0.06}
        \\
        & BK
        & 0.97
        & 0.49
        & { 0.06}
        & { 0.06}
        \\
        \hline
        \multirow{3}{*}{\shortstack{H$_2$O\\(-83.6)\,\, (14)}}
        & JW
        & 1.67
        & 0.51
        & 0.12
        & { 0.11}
        \\
        & Parity
        & 2.53
        & 0.65
        & 0.22
        & { 0.11}
        \\
        & BK
        & 3.26
        & 1.17
        & 0.20
        & { 0.10}
        \\
        \hline
        \multirow{3}{*}{\shortstack{NH$_3$\\(-66.9)\,\, (16)}}
        & JW
        & 3.79
        & 0.59
        & 0.18
        & 0.13
        \\
        & Parity
        & 5.22
        & 0.79
        & 0.21
        & 0.14
        \\
        & BK
        & 1.46
        & 0.61
        & 0.12
        & 0.11
        \\
        \hline
\end{tabular}
\end{table}

\subsection*{Acknowledgements}
I thank Antonio Mezzacapo for conversations about run-time complexity; Tim Laux and Thibault de Poyferr\'e for conversations about optimisation; Guglielmo Mazzola and Eric Peterson for comments on the manuscript; members of CSC for the stimulating environment while this work was being finished.

\end{section}

\begin{section}{Algorithm of Adaptive Pauli Shadows}

Consider a Hamiltonian $H$ on $n$ qubits. We write Pauli operators as $P=\otimes_{i\in[n]}P_i$ where $P_i\in\{I,X,Y,Z\}$. Then $H=\sum_P \alpha_P P$ where $\alpha_P\in\RR$. We shall say $P\in H$ if $\alpha_P\neq 0$. Set $\Bcal=\{X,Y,Z\}$ and consider a measurement basis $B=\otimes_{i\in[n]} B_i$ where $B_i\in\Bcal$. Measuring in the basis $B$ means measuring the $i$\textsuperscript{th} qubit in the basis $B_i$. If we prepare a state $\rho$ on a quantum processor of size $n$ qubits, and subsequently measure the quantum processor in the basis $B$, then we can estimate some Pauli observables $\Tr(P\rho)$. Specifically, we say $P$ is covered by $B$ if $P_i\in\{I, B_i\}$ for all $i\in[n]$. In the case $P$ is covered by $B$, a non-zero estimate of $\Tr(P\rho)$ may be given. 

Any algorithm which uses random measurements in Pauli bases to evaluate the energy of a Hamiltonian, given access to $\rho$, looks more or less like Algorithm~\ref{alg:energy}. The only difference is in the algorithm used to choose the measurement $B$ during the $s$\textsuperscript{th} sample.
(In the algorithm, $\mu_P$ is the best current estimate of $\Tr(P\rho)$ and $s_P$ is the number of times a basis has been chosen which covers $P$.)

\begin{algorithm}[h]
	\caption{Estimation of Energy via Adaptive Pauli Shadows}
	\label{alg:energy}
	\begin{algorithmic}
	    \State Initialise dictionary $D$ with keys $P\in H$ values $(\mu_P,s_P) \leftarrow (0,0)$
		\For{sample $s \in [S]$}
		    \State Set $B = \otimes_{i\in[n]} B_i$ by calling Algorithm~\ref{alg:basis}
		    \State Prepare $\rho$
			\For{$i\in[n]$}
			    \State Measure $i$\textsuperscript{th} qubit in basis $B_i$
			    \State Save eigenvalue measurement as $\sigma_i\in\{\pm 1\}$
			\EndFor
			\For{keys $P$ in dictionary $D$}
			    \If{$P$ is covered by $B$}
			        \State Update $\mu_P \leftarrow \frac{1}{s_P + 1}(s_P \mu_P + \prod_{i|P_i\neq I} \sigma_i)$
			        \State Update $s_P \leftarrow s_P + 1$
			    \EndIf
			\EndFor
		\EndFor
		\Return $\sum_P \alpha_P \mu_P$.
	\end{algorithmic}
\end{algorithm}

\newpage
Our goal is to describe Algorithm~\ref{alg:basis} and the computational problem which occurs.
On each qubit, we will build a probability distribution over the possible Pauli measurement bases $\Bcal = \{X, Y, Z\}$. This distribution will depend on the measurement bases already assigned to previous qubits. In order to reduce the importance of the ordering of qubits, we shall start by randomly choosing a bijection $i:[n]\to[n]$. Hence at stage $j\in[n]$ we will be considering qubit $i(j)$. Now, suppose that we are at stage $j\in[n]$ and we have assigned measurement bases $B_{i(j')}$ to qubits $i(j')$ for $j'<j$. Our goal is to build a probability distribution $\beta_{i(j)}:\Bcal\to\RR^+$ from which we shall randomly sample a measurement basis $B_{i(j)}$.
For expositional clarity, we shall simply write $\beta$ rather than $\beta_{i(j)}$.
Let us first define the set
\begin{equation}
    \Omega = \{ P \in H |   P_{i(j)}\in \Bcal 
                            \textrm{ and } 
                            P_{i(j')} \in \{I, B_{i(j')}\} 
                            \textrm{ whenever } j'<j \}.
\end{equation}
We now build the probability distribution $\beta:\Bcal\to\RR^+$ by solving the optimisation problem
\begin{equation}
\label{eqn:optimisation}
\begin{split}
    \textrm{minimise: }
    &
    \sum_{P\in \Omega} \frac{\alpha_P^2}{\beta(P_{i(j)})}
    \\
    \textrm{subject to: }
    &
    0\le \beta(B) \le 1 \textrm{ for } B\in\Bcal
    \\
    &
    \beta(X) + \beta(Y) + \beta(Z) = 1
\end{split}
\end{equation}
This problem is convex. In fact, there is an easy analytical solution which we shall describe. We can write $\Omega$ as a disjoint union $\Omega_X \cup \Omega_Y \cup \Omega_Z$ where $\Omega_B = \{ P\in \Omega | P_{i(j)}=B \}$. Writing $c_B = \sum_{P\in \Omega_B} \alpha_P^2$ the function to minimise reads simply $\sum_{B\in\Bcal} c_B/\beta(B)$. Lagrange multipliers provide the solution. Specifically, if $\sum_{B\in\Bcal}c_B = 0$ then we may set $\beta$ to be the uniform distribution. If $\sum_{B\in\Bcal}c_B>0$ then the solution is
\begin{equation}
    \beta(B) = \frac{\sqrt{c_B}}{\sum_{B'} \sqrt{c_{B'}}}.
\end{equation}

\begin{algorithm}
	\caption{Choice of measurement basis for Adaptive Pauli Shadows}
	\label{alg:basis}
	\begin{algorithmic}
	    \State Randomly choose a bijection $i:[n]\to[n]$
	    \For{$j\in[n]$}
	        \State Set $\beta_{i(j)}:\Bcal\to\RR^+$ by solving the optimisation problem in Eq.~\eqref{eqn:optimisation}
	        \State Choose $B_{i(j)}$ randomly according to distribution $\beta_{i(j)}$
	    \EndFor
	    \Return $B = \otimes_{i\in[n]} B_i$.
	\end{algorithmic}
\end{algorithm}

It remains to comment on the origin of the minimisation problem. To this end, we appeal to the ``diagonal cost function" studied in \cite[Eqn. (16)]{hadfield2020}. That convex minimisation problem allows us to establish a product probability distribution $\beta=\prod_{i\in[n]} \beta_i:\Bcal^n\to\RR^+$ over the set of Pauli measurement bases $\Bcal^n$. That is, when choosing a measurement basis $B=\otimes_{i\in[n]}B_i$, the choice on the $i$\textsuperscript{th} qubit is independent of other qubits. Our present minimisation problem is the obvious adaption of this cost function to the setting where we use knowledge of the measurement bases chosen on preceeding qubits (that is, on qubits $i(j')$ where $j'<j$).

Note that a simple run-time analysis of the optimisation problem in Eq.~\eqref{eqn:optimisation} gives a complexity $O(n_H \cdot n)$ where $n_H$ is the number of Pauli operators appearing in $H$ (that is, $P\in H$ with $\alpha_P\neq0$). This is better than derandomised classical shadows \cite{hkp2021} whose choice of $B_i$ also scales linearly with the number of preparations $S$ of the quantum state. (Moreover, derandomised classical shadows requests that the number of shots $S$ be declared before beginning the estimation procedure and also requires certain hyperparameters to be fixed beforehand. See \cite[Appendix C]{hkp2020}.)
This run-time complexity is in comparison to LBCS which is $O(1)$.

\end{section}



\bibliographystyle{unsrt}
\bibliography{references}

\end{document}